%% file: main.tex
\documentclass[
    aps,
    prab,
    reprint,
    superscriptaddress,
]{revtex4-2}

\makeatletter
\newif\ifispreprint
\@ifclasswith{revtex4-2}{preprint}{\ispreprinttrue}{\ispreprintfalse}
\makeatother

\ifispreprint
    \newcommand{\figwidth}{0.6\columnwidth}
\else
    \newcommand{\figwidth}{\columnwidth}
\fi

\usepackage{amsmath}
\usepackage{bm}
\usepackage{siunitx}

\usepackage{graphicx}

\usepackage[sort&compress]{natbib}

\usepackage{hyperref}
\hypersetup{
	colorlinks=true, breaklinks=true,
	linkcolor=black, citecolor=black, urlcolor=blue
}

\input{mathdef.tex}

\begin{document}

\title{Longitudinal beam instability driven by coherent radiation in an SSMB laser modulator}

\author{Yingjie Dai}
\author{Zhuoyuan Liu}
\author{Tong Li}
\affiliation{Department of Engineering Physics, Tsinghua University, Beijing 100084, China}
\author{Xiujie Deng}
\email{dengxiujie@mail.tsinghua.edu.cn}
\affiliation{Institute for Advanced Study, Tsinghua University, Beijing 100084, China}
\author{Lixin Yan}
\email{yanlx@mail.tsinghua.edu.cn}
\affiliation{Department of Engineering Physics, Tsinghua University, Beijing 100084, China}
\date{\today}

\begin{abstract}
Storage ring-based steady-state microbunching (SSMB) is a promising approach for generating high-average-power coherent radiation, while the instabilities driven by coherent undulator radiation in the laser modulator (LM) is important for the ring performance.
In this paper we investigate the longitudinal single-bunch multi-turn LM instability using cavity mode decomposition techniques.
The evolution of the wakefield in the longitudinal beam dynamics equations are derived, and the instability growth rates are analyzed.
Numerical simulations show excellent agreement with the theoretical model, validating the mode decomposition approach. These findings provide critical insights into the design and operation of SSMB storage rings, suggesting effective mitigation strategies to suppress the instability and enhance the overall performance.
\end{abstract}

\maketitle

\section{Introduction}

Advanced accelerator-based light sources have revolutionized various fields of science and technology by providing high-intensity, coherent radiation across a broad spectral range.
Currently, these sources are categorized into two main types: free electron lasers (FELs) and synchrotron radiation sources.
FELs utilize high-energy electron beams passing through undulators to generate high-peak-power coherent radiation, while synchrotron radiation sources rely on the bending of electron beams in storage rings to produce X-rays with high repetition rates.
The ongoing evolution of these facilities is driven by the persistent demand for higher brightness, shorter pulses, and improved coherence.

In 2010, Ratner and Chao proposed the concept of steady-state microbunching (SSMB) to combine the advantages of both FELs and synchrotron radiation
\cite{Ratner.2010.PRL.105.154801}.
In an SSMB storage ring, the conventional radiofrequency (RF) cavity is replaced by a laser modulator (LM) system, enabling microbunching at optical wavelengths.
The resulting sub-micrometer electron bunches capitalize on the high repetition rate of storage rings to produce intense, high-average-power radiation at short wavelengths.
Driven by this potential, SSMB has witnessed remarkable advancements over the past decade
\cite{Deng.2021.Nature.590.576,Deng.2020.PRAB.23.044001,Deng.2020.PRAB.23.044002,Kruschinski.2024.ComPhys.7.160,Deng.2021.PRAB.24.094001,Deng.2021.NIMA.1019.165859,Li.2023.PRAB.26.110701,Zhang.2021.PRAB.24.090701,Liu.2024.RSI.95.103004,Deng.2025.NST,Pan.2025.PRAB.28.114210,Chen.2026.PRAB.29.034602,Tang.2026.PRAB.29.024202}.

\begin{figure*}[t]
    \centering
    \includegraphics[width=0.7\textwidth, page=1]{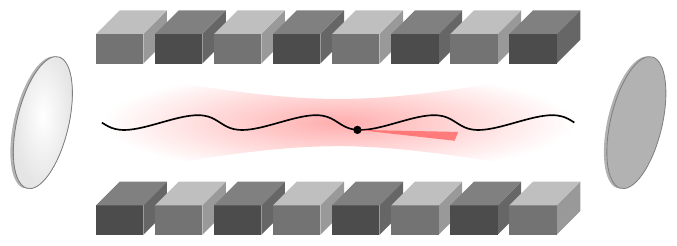}
    \caption{schematic of the laser modulator in an SSMB storage ring. The electron beam traverses the undulator with a periodic magnetic field, emitting radiation that bounces and evolves within the cavity. The interaction between the beam and the radiation field may trigger longitudinal beam instabilities.}
    \label{fig:laser modulator}
\end{figure*}

The LM system in an SSMB storage ring generally comprises an undulator and a laser cavity, as schematically illustrated in \figref{fig:laser modulator}.
When the electron beam traverses the undulator with a periodic magnetic field, it oscillates and emits radiation that will bounce and evolve within the cavity.
The interaction between the beam and the bouncing radiation field may trigger longitudinal beam instabilities, potentially degrading the beam quality and the overall performance of the storage ring.

In 2021, Tsai \textit{et al.} conducted pioneering and in-depth research on collective beam instabilities potentially driven by coherent radiation in an SSMB laser modulator
\cite{Tsai.2021.PRAB.24.114401,Tsai.2022.NIMA.167454,Tsai.2022.PRAB.25.064401,Tsai.2025.PRAB.28.074402,Tsai.2026.NIMA.171127}, employing an undulator radiation wake function combined with a macroparticle model.
To account for the recirculating nature of radiation within the optical cavity, the the radiation profile is assumed to be maintained over round-trips within the cavity, thereby enabling the description of multi-turn beam dynamics.
This assumption, in our view, may not strictly hold when considering the optical properties of the cavity, such as diffraction and mirror-induced focusing, which suggest a complex evolution of the field profile that warrants a detailed treatment.
To properly capture these phenomena, here we adopt a mode-decomposition approach.
By expanding the radiation into a complete set of cavity eigenmodes, we are able to represent the global evolution of the radiation as the sum of individual mode components.
Initial efforts to apply mode decomposition to radiation can be found in several prior studies
\cite{Kogelnik.1990.waveguides,Neuman.2001.PhD,Williams.2005.master,Vigil.2006.master,Gover.2005.PRAB.8.030701,Gover.2005.PRAB.8.030702,Niles.2010.PRAB.13.030702}.
This approach complements existing models by incorporating the detailed optical evolution of the modulator, providing further insights into the instability characteristics of SSMB storage rings.

In this paper, we first derive the expressions for the cavity modes and their excitation by the electron beam in section 2.
Then we decompose the undulator radiation spectrum into mode components and give the evolution of wakefields in the cavity in section 3.
Next, we formulate the longitudinal beam dynamics equations incorporating the radiation field and analyze the instability growth rates in section 4.
Finally, we present simulation results in section 5 and summarize our work in section 6.

\section{Cavity modes}

\subsection{Hermite--Gaussian modes}

In the optical cavity, the envelope $u(\bm{r})$ of a laser field with frequency $\omega$ propagating along the $z$-axis satisfies the paraxial wave equation
\cite{Siegman.1986.lasers}
\begin{equation}
    \label{eqn:paraxial wave equation}
    \nabla_\perp^2u - \i 2k\pv uz = 0,
\end{equation}
where $\nabla_\perp^2$ is the transverse Laplacian and $k = \omega / c$ is the wavenumber.
The convention $\bm{E}(\bm{r}, t) = \bm{E}(\bm{r}) \e^{\i\omega t}$ and $E(\bm{r}) = u(\bm{r}) \e^{-\i kz}$ are used.
In Cartesian coordinates, the solutions of \eqref{eqn:paraxial wave equation} are the Hermite--Gaussian modes, also known as the transverse electromagnetic (TEM) modes:
\begin{widetext}
\begin{equation}
    \label{eqn:HG mode}
    \begin{split}
        E_{mn} = E_0&\frac{w_0}{w(z)} H_m\biggp{\frac{\sqrt{2}x}{w(z)}} H_n\biggp{\frac{\sqrt{2}y}{w(z)}} \exp\biggp{-\frac{x^2 + y^2}{w^2(z)}}\\
        &\times \exp\biggb{-\i\frac{k(x^2+y^2)}{2R(z)} + \i(m+n+1)\arctan\biggp{\frac{z}{\zR}}-\i kz},
    \end{split}
\end{equation}
\end{widetext}
where $E_0$ is the amplitude normalization constant, $H_m$ and $H_n$ are Hermite polynomials of order $m$ and $n$ respectively, $w_0=w(0)$ is the beam waist $w(z)$ at $z=0$.
The beam waist $w(z)$, radius of curvature $R(z)$, and Rayleigh length $\zR$ are defined as:
\begin{align}
    w(z) &= w_0 \sqrt{1 + \Bigp{\frac{z}{\zR}}^2}, \\
    R(z) &= z + \frac{\zR^2}{z}, \\
    \zR &= \frac{\pi w_0^2}{\lambda}.
\end{align}
The validity of the paraxial approximation requires that the divergence angle $\Theta = w_0/\zR \ll 1$, i.e.
\begin{equation}
    \label{eqn:paraxial approximation}
    k\zR \gg 1,
\end{equation}
which is equivalent to the condition that the envelope $u(\bm{r})$ varies slowly along the $z$-direction compared to the wavelength $\lambda$.

\begin{figure}[t]
    \centering
    \includegraphics[width=\figwidth]{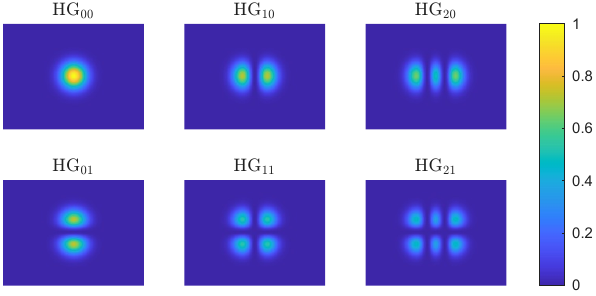}
    \caption{Intensity profiles of Hermite--Gaussian modes with different mode orders $(m,n)$ at $z=0$.
    The intensity is normalized to the total power of the mode.}
    \label{fig:mode intensity}
\end{figure}

The intensity profile $I\propto\abs{E_{mn}}^2$ is given by the product of a Gaussian envelope and the squared Hermite polynomials, which has $(m+1)$ and $(n+1)$ peaks along the $x$- and $y$-directions respectively, as shown in \figref{fig:mode intensity}.

\subsection{Phase contour and resonance condition}

For mode confinement in a laser cavity formed by two reflective mirrors, the mirror profiles need to conform to the wavefront of the modes at the respective mirror positions.
The phase of the mode is
\begin{equation}
    \Phi = - \frac{k(x^2+y^2)}{2R(z)} + (m+n+1)\arctan\biggp{\frac{z}{\zR}} -kz + \omega t,
\end{equation}
the radius of curvature of the phase contour at position $z$ is
\begin{equation}
    \rho = z + \frac{\zR^2}{z} - (m+n+1)\frac{\zR}{k z}.
\end{equation}
The paraxial approximation \eqref{eqn:paraxial approximation} implies that the last term is negligible, so $\rho \approx R(z) = z + \zR^2/z$.
Typically, the cavity is designed for the fundamental mode (HG$_{00}$) at a target laser wavelength $\lambda_\las$ with a specific Rayleigh length $\zRlas$.
The geometric boundary conditions of the cavity then dictate the Rayleigh length $\zR(\lambda)$ and beam waist $w_0(\lambda)$ for other modes and wavelengths.
Over a broad spectral range around $\lambda_\las$, the Rayleigh length can be approximated as constant $\zR(\lambda) \approx \zRlas$, which remains highly accurate under the paraxial approximation.

In the laser modulator, the laser beam is focused to modulate the electron beam in the undulator.
Optimization shows that the effective modulation voltage $V_\las$ peaks at
\cite{Deng.2024.PhD}
\begin{equation}
    \frac{\zR}{L_\und} = \arg\max_z \sqrt{z} \arctan\Bigp{\frac{1}{2z}} \approx 0.359,
\end{equation}
and $V_\las$ remains relatively insensitive to $\zR$ for values exceeding this maximum point.
Consequently, in practical designs, the Rayleigh length $\zR$ is usually chosen to be on the order of the undulator length $L_\und$, which means the paraxial approximation is well satisfied for the laser modulator.

Furthermore, to support stable modes within the cavity, the resonance condition must be satisfied. This requires the round-trip phase shift of the modes to be an integer $q = 1,2,3,\ldots$ multiple of $2\pi$, leading to the formation of a standing wave pattern. For a two-mirror cavity with identical radii of curvature $R_1 = R_2 = R$, the resonant frequencies are given by
\cite{Kogelnik.1966.laser}
\begin{equation}
    \label{eqn:resonance frequency}
    \omega_{mnq} = \frac{c}{L}\Bigb{q\pi + (m + n + 1)\arccos\Bigp{1 - \frac{L}{R}}},
\end{equation}
where $L$ is the cavity length.
Since $L \gg \lambda_\las$, the free spectral range (FSR) of the modes
\begin{equation}
    \Delta\omega_\FSR = \omega_{mn(q+1)} - \omega_{mnq} = \frac{\pi c}{L}
\end{equation} 
is significantly smaller than the laser frequency $\omega_\las$, implying that the longitudinal mode index $q$ is typically very large.
Meanwhile, the transverse mode indices $m, n$ are relatively small, as higher-order modes exhibit larger beam sizes and suffer from higher diffraction losses.
Nevertheless, the transverse phase shift $\psi_{mn} = (m + n + 1)\arccos(1 - L/R)$ introduced by Gouy phase plays a crucial role in the mode evolution within the cavity, distinguishing it from free-space propagation. This will be discussed in detail in the next section.

\subsection{Mode expansion and excitation}

However, such Hermite--Gaussian modes have only transverse electric field components, which does not satisfy the Maxwell's equations.
To be consistent with Maxwell's equations,
the longitudinal electric field component $E_z$ needs to be derived from the divergence-free condition of the electric field:
\begin{equation}
    \label{eqn:divE=0}
    \nabla \cdot \bm{E} = \pv{E_x}{x} + \pv{E_y}{y} + \pv{E_z}{z} = 0.
\end{equation}
As $E_{mn}(x,y,z) = E_{nm}(y,x,z)$, without loss of generality, we here consider the $x$-polarization with $E_x = E_{mn}$ and $E_y = 0$.
Using the slowly varying envelope approximation, $\p E_z/\p z\approx -\i k E_z$, then \eqref{eqn:divE=0} gives:
\begin{equation}
    \label{eqn:Ez}
    E_z \approx -\frac{\i}{k} \pv{E_x}{x},
\end{equation}
and we get the expression of $\bm E_{mn}$.
Then the magnetic field $\bm H_{mn}$ can be derived from Faraday's law:
\begin{equation}
    \bm H = \frac{\i}{\omega \mu_0}\nabla\times\bm E.
\end{equation}

For the completeness of Hermite polynomials,
we can express any paraxial field in terms of Hermite--Gaussian modes:
\begin{subequations}
    \begin{align}
        \bm{E}(\bm{r}) &= \sum_{mn} C_{mn}\bm{E}_{mn}(\bm{r}), \\
        \bm{H}(\bm{r}) &= \sum_{mn} C_{mn}\bm{H}_{mn}(\bm{r}).
    \end{align}
\end{subequations}
then the evolution of the field can be described by the evolution of the mode amplitudes $C_{mn}$.
Note that field $\bm E(\bm r)$ is the Fourier transforms at frequency $\omega$ of the $\bm E(\bm r,t)=\bm{E}(\bm{r})\e^{\i\omega t}$ in time-domain, thus $C_{mn}$ has the dimension of time.

The excitation of cavity modes by current density $\bm{J}$ can be described using the mode amplitude equation
\cite{Gover.2005.PRAB.8.030701}
\begin{equation}
    \label{eqn:mode excitation general}
    \Delta C_{mn} = -\frac{1}{4P_{mn}} \int \bm{J}(\bm{r}) \cdot \conj{\bm{E}_{mn}}(\bm{r})\d V,
\end{equation}
where $P_{mn}$ is the power normalization constant of the mode:
\begin{equation}
    P_{mn} = \frac{1}{2} \int \Re[\bm{E}_{mn}(\bm{r}) \times \conj{\bm{H}_{mn}}(\bm{r})] \cdot \hat{\bm z} \d A.
\end{equation}
for the Hermite--Gaussian modes, it can be shown that:
\begin{equation}
    \label{eqn:mode power}
    P_{mn} = \frac{1}{4}\varepsilon_0E_0^2\cdot\pi w_0^2\cdot c\cdot(2m)!!(2n)!!,
\end{equation}
where $E_0$ is the amplitude normalization constant of the modes in \eqref{eqn:HG mode}.
This expression admits a straightforward physical interpretation: it is the product of the energy density ($\varepsilon_0E_0^2/4$), the laser beam cross-section area ($\pi w_0^2$), and the speed of light ($c$).
The double factorial factor $(2m)!!(2n)!!$ arise from the inner product of Hermite polynomials.

The energy spectrum of the radiation in terms of the mode amplitudes is given by the Parseval theorem:
\begin{equation}
    \dv{W_{mn}}{\omega} = \frac{2}{\pi} P_{mn} \abs{\Delta C_{mn}}^2.
\end{equation}
and the total radiation energy spectrum is given by summing over all modes:
\begin{equation}
    \dv{W}{\omega} = \sum_{mn} \dv{W_{mn}}{\omega} = \frac{2}{\pi} \sum_{mn} P_{mn} \abs{\Delta C_{mn}}^2.
\end{equation}

If the current density is given by a series of electrons at positions $\bm{r}_{i}(t)$ with velocities $\bm{v}_{i}(t)$:
\begin{equation}
    \bm{J}(\bm{r},t)=-e\sum_{i=1}^{N_\elec} \bm{v}_{i}(t) \delta\bigp{\bm{r}-\bm{r}_{i}(t)}.
\end{equation}
Substituting into \eqref{eqn:mode excitation general} and taking the Fourier transform of $\bm{J}(\bm{r},t)$ gives
\cite{Gover.2005.PRAB.8.030701}
\begin{subequations}
    \label{eqn:mode excitation}
    \begin{align}
        \Delta C_{mn} &= -\frac{1}{4P_{mn}} \sum_{i=1}^{N_\elec} \Delta W_{mn,i},\\
        \Delta W_{mn,i} &= -e \int\iti \bm{v}_{i}(t) \cdot \conj{\bm{E}_{mn}}\bigp{\bm{r}_{i}(t), t}\d t,
    \end{align}
\end{subequations}
where $\Delta W_{mn,i}$ is the work done by $i$-th electron on the mode field $\bm{E}_{mn}$ over its trajectory $\bm{r}_{i}(t)$

Next, we will use this equation to calculate the modes of the radiation excited by the electron beam in the laser modulator.

\section{Mode decomposition of undulator radiation}

The prescribed motion of a relativistic electron with an energy $\gamma m_\elec c^2$ in a planar undulator is given by
\cite{Chao.2020.Lectures}
\begin{subequations}
    \label{eqn:undulator motion}
    \begin{align}
        v_x&=-c\frac{K}{\gamma}\sin(k_\und\bar v_zt),\\
        v_y&=0,\\
        v_z&=\bar v_z+c\frac{K^2}{4\gamma^2}\cos(2k_\und\bar v_zt),\\
        x&=\frac{K}{\gamma k_\und}\cos(k_\und\bar v_zt),\\
        y&=0,\\
        z&=\bar v_zt+\frac{K^2}{8\gamma^2k_\und}\sin(2k_\und\bar v_zt),
    \end{align}
\end{subequations}
where $K$ is the dimensionless undulator parameter, $k_\und=2\pi/\lambda_\und$ is the undulator wavenumber, $\bar v_z$ is the average longitudinal velocity.
\begin{equation}
    \bar v_z = c\biggp{1-\frac{1+K^2/2}{2\gamma^2}}.
\end{equation}

\subsection{Undulator radiation spectrum}

The double differential spectrum of a moving electron can be derived from the Li\'enard--Wiechert potential
\cite{Jackson.2021.electrodynamics}
\begin{equation}
    \dw{W}{\omega}{\Omega} = \frac{e^2\omega^2}{16\pi^3\varepsilon_0c^3} \biggabs{\int\iti \hat{\bm n} \times (\hat{\bm n} \times \bm v) \e^{-\i\omega t + \i k\hat{\bm n}\cdot\bm r}\d t}^2,
\end{equation}
Substituting the electron motion from \eqref{eqn:undulator motion} into the above equation, we can get the well-known undulator radiation spectrum
\cite{Chao.2020.Lectures}
\begin{equation}
    \dw{W_H}{\omega}{\Omega} = \frac{2e^2\gamma^2}{\pi\varepsilon_0c}F(\epsilon)G(\theta,\phi),
\end{equation}
the radiation coherents at discrete harmonics $H=1,2,3,\ldots$ of the fundamental resonant wavelength $\lambda_\rad$,
\begin{equation}
    \lambda_\rad=\frac{\lambda_\und}{2\gamma^2}\biggp{1+\frac{K^2}{2}+\gamma^2\theta^2},
\end{equation}
factor $F(\epsilon)$ describes the frequency dependence around the harmonic,
\begin{equation}
    F(\epsilon) = \biggp{\frac{\sin(N_\und\pi\epsilon)}{\pi\epsilon}}^2,
\end{equation}
where $\epsilon = \omega/\omega_\rad - H$ is the relative frequency detuning from the $H$-th harmonic, $N_\und$ is the number of undulator periods.
While factor $G(\theta, \phi) = G_\sigma(\theta, \phi) + G_\pi(\theta, \phi)$ describes the angular distribution of the radiation, the $\sigma$- and $\pi$-mode polarization correspond to the $x$- and $y$-component of $\hat{\bm n}\times(\hat{\bm n}\times\bm v)$ respectively,
\begin{subequations}
    \begin{align}
        G_\sigma(\theta,\phi) &= \frac12\biggp{\frac{H(K\mathcal{D}_1+\gamma\theta\mathcal{D}_2\cos\phi)}{1+K^2/2+\gamma^2\theta^2}}^2,\\
        G_\pi(\theta,\phi) &= \frac12\biggp{\frac{H\gamma\theta\mathcal{D}_2\sin\phi}{1+K^2/2+\gamma^2\theta^2}}^2,
    \end{align}
\end{subequations}
where $\theta$ and $\phi$ are the polar and azimuthal angles of the observation direction $\hat{\bm n}$, and $\mathcal{D}_1,\mathcal{D}_2$ are given by series of Bessel functions,
\begin{subequations}
    \begin{align}
        \mathcal{D}_1 &= -\frac12\sum_{m=-\infty}^\infty J_{H+2m-1}(H\alpha) \bigb{J_m(H\zeta)+J_{m-1}(H\zeta)},\\
        \mathcal{D}_2 &= \sum_{m=-\infty}^\infty J_{H+2m}(H\alpha)J_m(H\zeta),
    \end{align}
\end{subequations}
where $\alpha,\zeta$ are defined as:
\begin{subequations}
    \begin{align}
        \alpha &= \frac{2K\gamma\theta\cos\phi}{1+K^2/2+\gamma^2\theta^2},\\
        \zeta &= \frac{K^2/4}{1+K^2/2+\gamma^2\theta^2}.
    \end{align}
\end{subequations}

\begin{figure}[t]
    \centering
    \includegraphics[width=\figwidth]{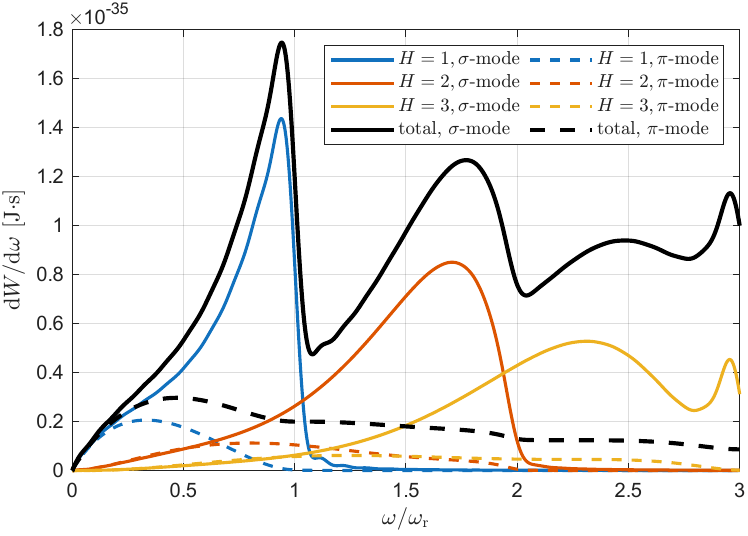}
    \caption{Undulator radiation spectrum.
    The blue, orange, yellow, and black lines are for the fundamental ($H=1$), second ($H=2$), third ($H=3$), and total (summing over $H=1,2,3,\ldots$) harmonics respectively, with the solid lines for $\sigma$-mode and the dashed lines for $\pi$-mode.}
    \label{fig:spectrum H}
\end{figure}

The frequency spectrum of the radiation can be obtained by integrating the double differential spectrum over the solid angle $\Omega$:
\begin{equation}
    \dv{W_H}{\omega}=\oint \dw{W_H}{\omega}{\Omega}\d\Omega=\int_0^\pi \int_0^{2\pi} \dw{W_H}{\omega}{\Omega} \sin\theta\d\theta\nd\phi.
\end{equation}
By splitting factor $G(\theta, \phi)$ into $G_\sigma(\theta, \phi)$ and $G_\pi(\theta, \phi)$, we can also get the radiation spectrum of each polarization mode $\nd W_{H,\sigma}/\nd\omega$ and $\nd W_{H,\pi}/\nd\omega$ separately,
the undulator radiation spectrum of $\sigma$- and $\pi$-modes are shown in \figref{fig:spectrum H} with the parameters in \tabref{tab:parameters} to be presented below.
The total radiation spectrum is given by summing over all harmonics,
\begin{equation}
    \dv{W}{\omega}\biggr\vert_{\text{LW}} = \sum_H\dv{W_H}{\omega} = \sum_H\dv{W_{H,\sigma}}{\omega} + \sum_H\dv{W_{H,\pi}}{\omega}.
\end{equation}

\begin{figure}[t]
    \centering
    \includegraphics[width=\figwidth]{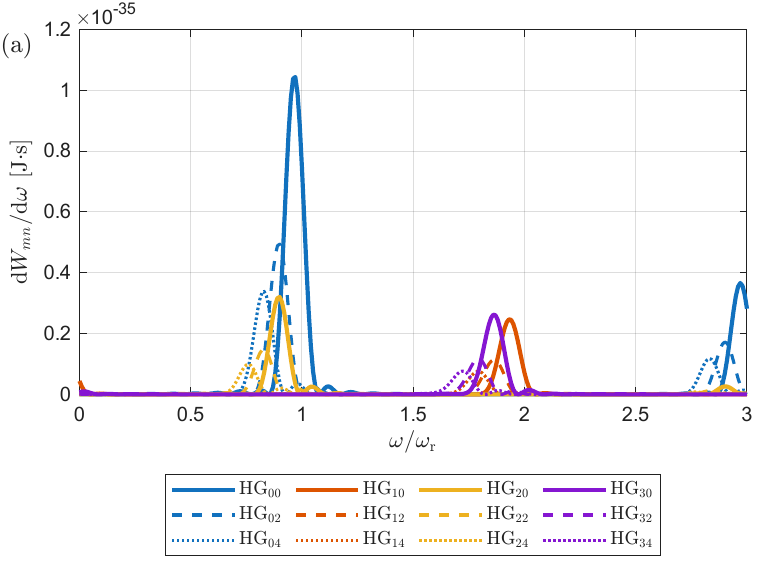}\\[2ex]
    \includegraphics[width=\figwidth]{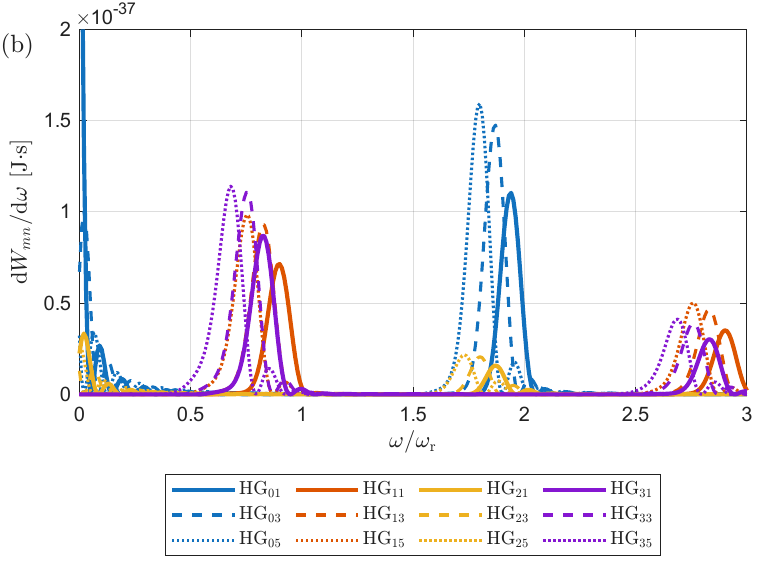}
    \caption{Hermite--Gaussian mode components of undulator radiation.
    Different line colors are for different $m$, and different line styles are for different $n$.
    (a) The $\sigma$-mode radiation has only even-$n$ modes, while (b) the $\pi$-mode radiation has only odd-$n$ modes.}
    \label{fig:undulator radiation modes}
\end{figure}

Using the mode excitation equation \eqref{eqn:mode excitation} and Parsval theorem, we can calculate the undulator radiation in terms of cavity modes.
\begin{equation}
    \dv{W_{mn}}{\omega} = \frac{e^2}{8\pi P_{mn}} \biggabs{\int\iti \bm{v} \cdot \conj{\bm{E}_{mn}}(\bm r, t)\d t}^2,
\end{equation}
as $P_{mn} \propto E_0^2$ and $\bm{E}_{mn} \propto E_0$, $\nd W_{mn}/\nd\omega$ is independent of the choice of $E_0$.
The integral interval is limited within the undulator length $-L_\und/2\bar v_z \leq t \leq L_\und/2\bar v_z$ as the electron does not emit significant radiation outside the undulator.

The $x$- and $y$-polarized components of the undulator radiation can be decomposed into Hermite--Gaussian modes, corresponding to $(E_x = E_{mn}, E_y = 0)$ and $(E_x = 0, E_y = E_{mn})$ respectively.
Due to the absence of electron motion along the $y$-axis, only the even-$n$ modes with $x$-polarization ($\sigma$-mode) and odd-$n$ modes with $y$-polarization ($\pi$-mode) are excited, and the $\pi$-mode radiation is much weaker than the $\sigma$-mode radiation.

The Hermite--Gaussian mode components of undulator radiation are shown in \figref{fig:undulator radiation modes} with the parameters in \tabref{tab:parameters}.
One can see that the peaks of the mode spectrum are around the harmonics of the fundamental resonant frequency $\omega_\rad$, with even-$(m + n)$ modes peaking at odd harmonics and odd-$(m + n)$ modes peaking at even harmonics.
While the fundamental mode (HG$_{00}$) is the strongest mode, high-order modes with $m, n > 0$ also contribute significantly to the total radiation spectrum, especially for higher harmonics.
The mode spectrum exhibits an unphysical (though non-divergent) rise when $\omega \to 0$.
Nevertheless, as the final results turn out, this discrepancy is localized to the low-frequency regime and does not compromise the accuracy of the mode decomposition near the harmonics, which are of primary interest for analyzing beam instabilities in the laser modulator.

\begin{figure}[t]
    \centering
    \includegraphics[width=\figwidth]{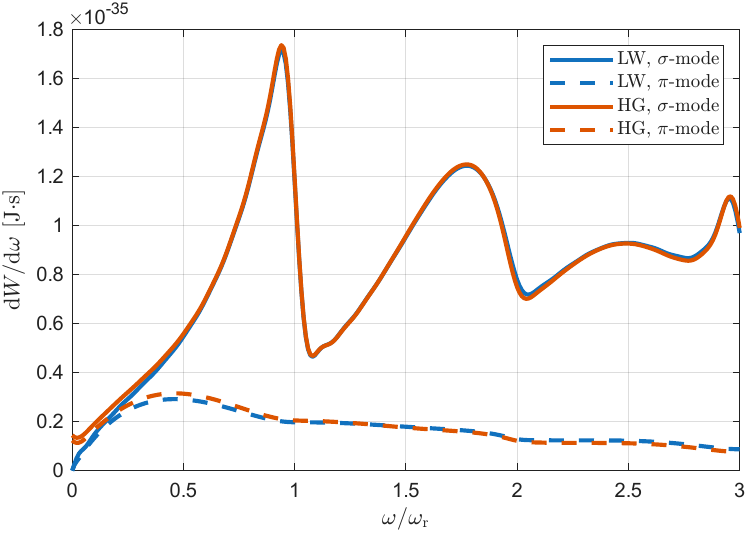}
    \caption{Comparison of undulator radiation spectrum obtained from the Li\'enard--Wiechert potential and the Hermite--Gaussian mode decomposition.
    The blue line is the spectrum obtained from the Li\'enard--Wiechert potential, and the orange line is the spectrum obtained from the Hermite--Gaussian mode decomposition.
    The solid lines are for the $\sigma$-mode and the dashed lines are for the $\pi$-mode.}
    \label{fig:spectrum comparison}
\end{figure}

The total radiation spectrum is given by summing over all modes,
\begin{equation}
    \dv{W}{\omega}\biggr\vert_{\text{HG}} = \sum_{mn} \dv{W_{mn}}{\omega}.
\end{equation}
As shown in \figref{fig:spectrum comparison}, the spectrum obtained via Hermite--Gaussian mode decomposition is in excellent agreement with the Liénard--Wiechert potential results, thereby validating the mode decomposition approach for analyzing undulator radiation.

\begin{figure}[t]
    \centering
    \includegraphics[width=\figwidth]{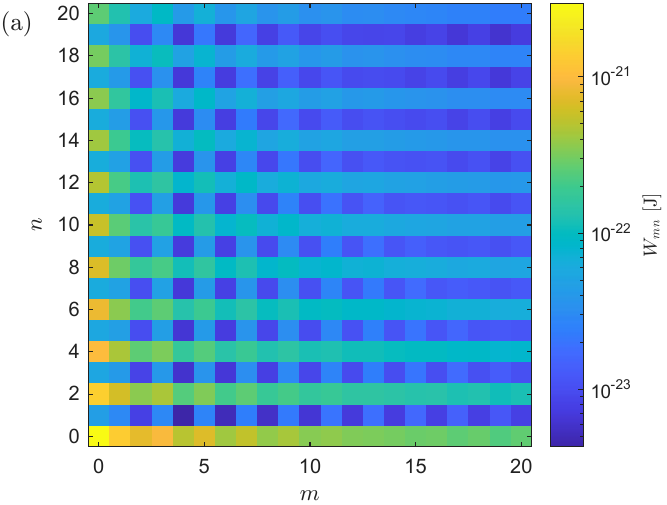}\\[2ex]
    \includegraphics[width=\figwidth]{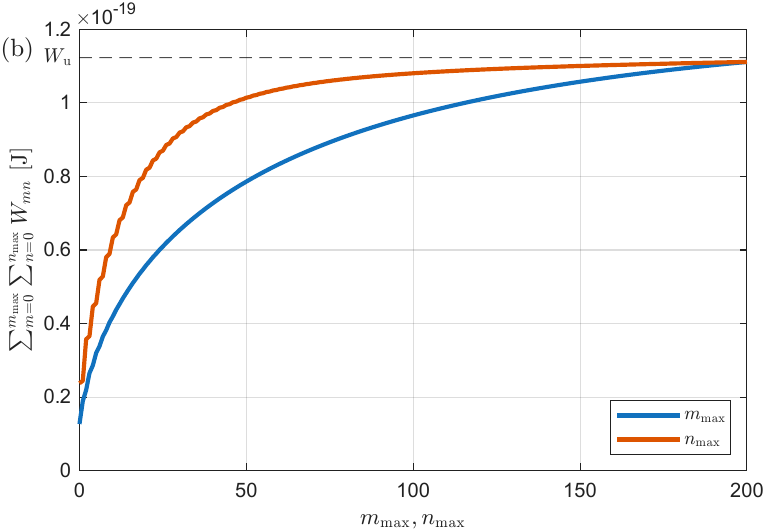}
    \caption{Convergence of the mode energy of the undulator radiation spectrum.
    (a) The energy $W_{mn}$ distribution of each mode with mode order $m, n \leq 20$.;
    (b) The blue line is the energy cumulatively summing over $m \leq m_{\max}$ and $n \leq 200$, the orange line is the energy cumulatively summing over $m \leq 200$ and $n \leq n_{\max}$.
    The dashed line is the theoretical total energy $W_\und = \SI{1.12e-19}{J}$ of the undulator radiation in \eqref{eqn:undulator radiation energy}.}
    \label{fig:spectrum mode convergence}
\end{figure}

The convergence of the mode decomposition is usually not very fast though, therefore we need to include enough high-order modes to get accurate results.
The radiation energy of each mode is obtained by integrating the spectrum over frequency:
\begin{equation}
    W_{mn} = \int\zti \dv{W_{mn}}{\omega} \d\omega.
\end{equation}
As shown in \figref{fig:spectrum mode convergence}(a), the distribution of the mode energy is related with the parity of $m, n$ because of the characteristics of the undulator motion and the mode intensity profile.
Adding up all modes $W_{mn}$ gives the total energy of the undulator radiation;
while the total energy of the undulator radiation can also be calculated by integrating the spectrum from Li\'enard--Wiechert potential
\cite{Chao.2020.Lectures},
\begin{equation}
    \label{eqn:undulator radiation energy}
    W_\und = \frac{e^2 \gamma^2 K^2 k_\und N_\und}{6\varepsilon_0}.
\end{equation}
As shown in \figref{fig:spectrum mode convergence}(b), the total energy obtained by summing over modes converges to the theoretical value $W_\und$.

\subsection{Wakefield and its evolution in the cavity}

The electromagnetic interaction between the beam and its environment is described by radiation impedance in the frequency domain and wakefields in the spatial domain.
The real part of the impedance, which accounts for the resistive energy dissipation, is directly proportional to the radiation spectrum for a single electron, as given by the following relation
\cite{Chao.1993.instability,Chao.2020.Lectures}
\begin{equation}
    \label{eqn:undulator radiation impedance}
    \Re[Z_{\parallel}(\omega)] = \frac{\pi}{e^2} \dv{W}{\omega},
\end{equation}
we can also define the impedance of each harmonic $\Re[Z_{\parallel,H}(\omega)]$ and each mode $\Re[Z_{\parallel,mn}(\omega)]$ accordingly.

While the impedance provides a frequency-domain representation of the interaction strength, the longitudinal wakefield function $W_{\parallel}(z)$ describes the normalized integrated radiation interaction by trailing particles on the leading ones at a distance $z$ due to the longitudinal velocity $v_z < c$.
The longitudinal wake function in free space can be reconstructed from the real part of the impedance via the inverse cosine transform $W_{\parallel}(z) = \IFT_{\cos}[Z_{\parallel}](z)$
\cite{Chao.1993.instability,Chao.2020.Lectures}
\begin{subequations}
    \begin{align}
        \label{eqn:wake in free space}
        \IFT_{\cos}[Z_{\parallel}](z) &:= \frac{2}{\pi} \int\zti \Re[Z_{\parallel}(\omega)] \cos(k z) \d\omega;\\
        \IFT_{\sin}[Z_{\parallel}](z) &:= \frac{2}{\pi} \int\zti \Re[Z_{\parallel}(\omega)] \sin(k z) \d\omega.
    \end{align}
\end{subequations}
Here the inverse sine transform $\IFT_{\sin}[Z_{\parallel}](z)$ is also defined for later use.

\begin{figure}[!t]
    \centering
    \includegraphics[width=\figwidth]{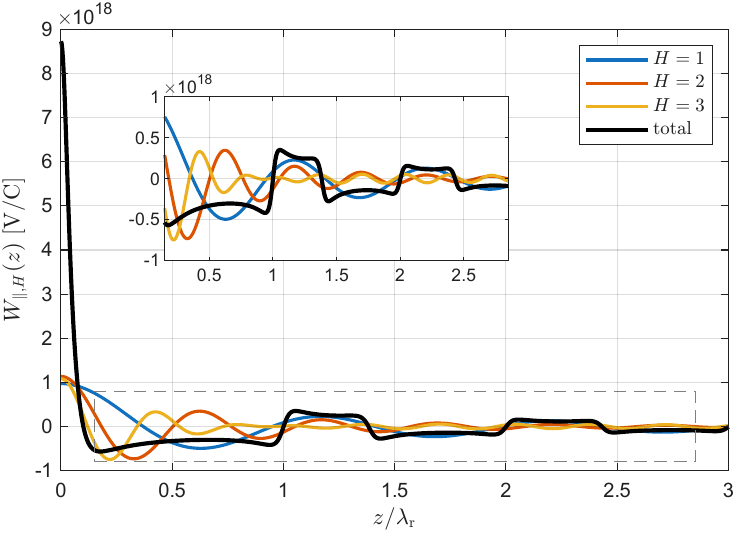}
    \caption{Longitudinal wakefield of undulator radiation for the first three harmonics and the total sum.
    The blue, orange, and yellow lines represent the fundamental ($H=1$), second ($H=2$), and third ($H=3$) harmonics, respectively.
    The thick black line denotes the total wakefield.}
    \label{fig:wake H}
\end{figure}

\figref{fig:wake H} illustrates the wakefield calculated using the parameters listed in \tabref{tab:parameters}. While the wakefield of individual harmonics exhibits a damped sinusoidal profile, their superposition results in a highly localized, short-range total wakefield that attenuates rapidly from its peak value $W_{\parallel}(0) = 2W_\und/e^2 = \SI{8.73e18}{V/C}$.

The frequency spectrum of the modes, however, is discrete due to the cavity resonance condition \eqref{eqn:resonance frequency}, thus the Fourier transform integral in the wake function $W_{\parallel}(z)$ should be replaced by Fourier series:
\begin{equation}
    \label{eqn:wake in cavity}
    W_{\parallel}(z) = \frac{2}{\pi} \sum_{mnq} \Re[Z_{\parallel, mn}(\omega_{mnq})]\cos(k_{mnq}z) \Delta\omega_\FSR,
\end{equation}
where $\Re[Z_{\parallel, mn}]$ is the the impedance of HG$_{mn}$ modes, $k_{mnq} = \omega_{mnq} / c = (\psi_{mn} + \pi q) / L$ is the wavenumber of the modes, $\psi_{mn} = (m + n + 1)\arccos(1 - L/R)$ is the transverse phase shift introduced by the Gouy phase, and $\Delta\omega_\FSR = \pi c / L$ is the the free spectral range of the modes.
Typically, the wavelength of the radiation $\lambda_\rad$ is much smaller than the cavity length $L$, thus the modes are highly dense in the frequency domain with $\Delta\omega_\FSR \ll \omega_\rad$.
Under this condition, the wakefield in the cavity has two distinct regimes where the characteristics are completely different: the short-range regime where $z \ll L$ and the long-range regime where $z$ approaches the cavity scale $L$.

In the short-range regime where $z \ll L$, the Euler--Maclaurin formula ensures that the discrete Fourier series of $W_{\parallel}(z)$ asymptotically approaches the continuous Fourier transform $\IFT_{\cos}[Z_{\parallel}](z)$ in \eqref{eqn:wake in free space}, the asymptotic behavior could be denoted as:
\begin{equation}
    \label{eqn:wake short range}
    W_{\parallel}(z) \simeq \IFT_{\cos}[Z_{\parallel}](z), \quad z \ll L.
\end{equation}
Physically, this implies that the short-range wakefield is dominated by the local radiation characteristics, and the influence of the cavity boundaries is negligible.

\begin{figure*}[t]
    \centering
    \includegraphics[width=\linewidth]{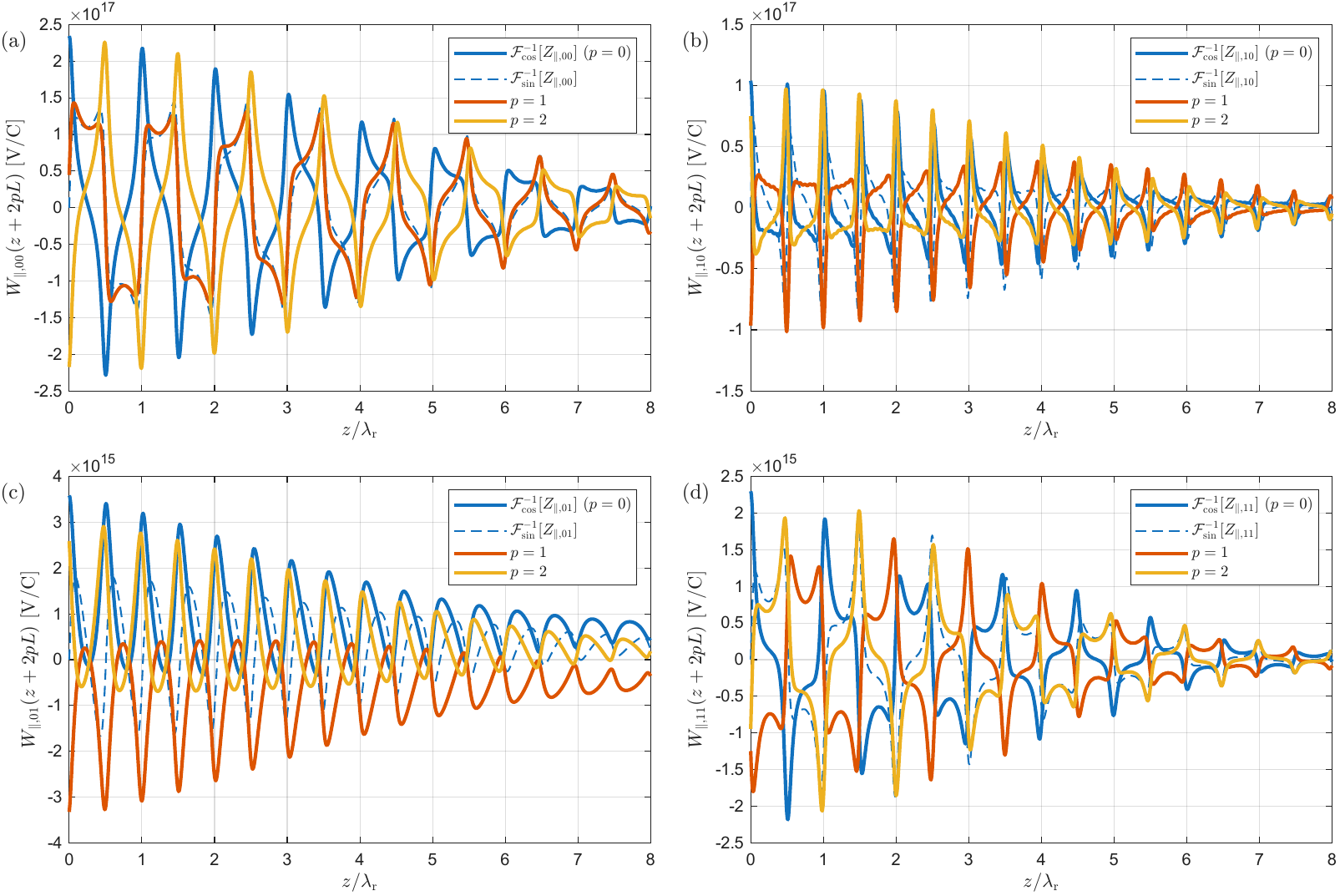}
    \caption{Evolution of the wakefield $W_{\parallel,mn}(2pL + z)$ of the (a) HG$_{00}$ mode, (b) HG$_{10}$ mode, (c) HG$_{01}$ mode, and (d) HG$_{11}$ mode over circulations.
    The blue, orange, and yellow solid lines represent the wakefield after zero ($p=0$), one ($p=1$), and two ($p=2$) round-trip reflections within the cavity, respectively.
    For reference, the inverse cosine transform $\IFT_{\cos}[Z_{\parallel,00}]$ (corresponding to the $p=0$ initial wake) and the inverse sine transform $\IFT_{\sin}[Z_{\parallel,00}]$ of the impedance are displayed as blue solid and dashed lines.
    Here the phase shift $\psi_{00} = \arccos(1 - L/R) \approx 2.45$ with the cavity parameters in \tabref{tab:parameters}.}
    \label{fig:wake modes}
\end{figure*}

In the long-range regime, conversely, as $z$ approaches the cavity scale $L$, the discreteness of the spectrum becomes manifest, leading to significant deviations from the short-range regime.
To clarify that, we can rewrite the expression of wake function \eqref{eqn:wake in cavity} by explicitly separating the Gouy phase term $\psi_{mn}$ and the Fourier series term:
\begin{equation}
    W_{\parallel}(z) = \frac{2c}{L} \sum_{mn} \Re\biggb{\e^{\i\psi_{mn} z/L} \sum_{q} \Re[Z_{\parallel, mn}(\omega_{mnq})]\e^{\i\pi q z/L}},
\end{equation}
which indicates that the wakefield in the cavity exhibits a quasi-periodic structure with a fundamental period of $2L$ driven by the Fourier series term $\sum_{q} \Re[Z_{\parallel, mn}(\omega_{mnq})]\e^{\i\pi q z/L}$, while the Gouy phase $\psi_{mn}$ introduces a cumulative phase slip that modulates the interference between transverse modes, so generally the wakefield does not have a strict periodicity of $2L$:
\begin{equation}
    W_{\parallel}(z + 2L) \neq W_{\parallel}(z).
\end{equation}

\begin{figure}[!t]
    \centering
    \includegraphics[width=\figwidth]{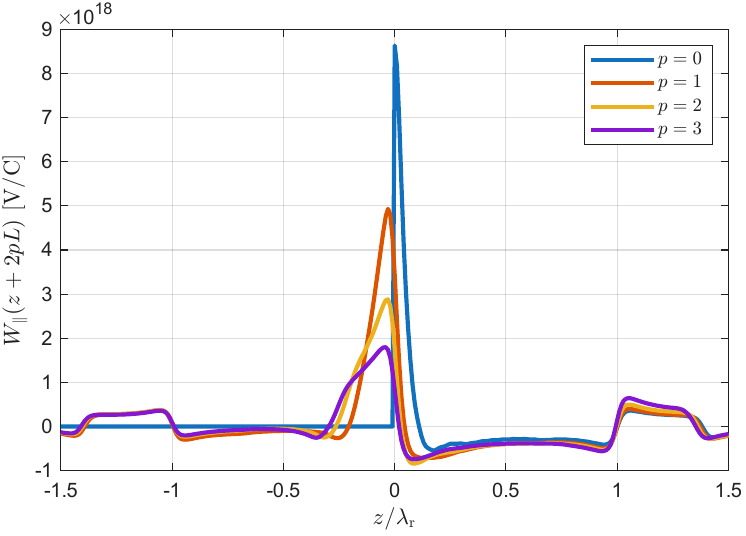}
    \caption{Evolution of the longitudinal wakefield $W_{\parallel}(z + 2pL)$.
    The blue, orange, yellow, and purple solid lines represent the wakefield after zero ($p=0$), one ($p=1$), two ($p=2$), and three ($p=3$) round-trip reflections within the cavity, respectively.
    Here the phase shift $\psi_{00}$ is set to 0.01 to show a small evolution.}
    \label{fig:wake evolution}
\end{figure}

The direct numerical calculation of the wake function \eqref{eqn:wake in cavity} is computationally expensive due to a large number of $q$ at the scale of $L / \lambda_\rad$.
However, we don't need to calculate the long-range wakefield $W_{\parallel}(z)$ across the entire cavity length $L$.
For SSMB applications, the electron bunch only samples the wakefield turn-by-turn around the same phase space bucket, while the wakefield evolves after several round-trips in the cavity.
Due to this periodic synchronization, our analysis can be restricted to the localized properties of the wakefield on multiple round-trips by shifting the longitudinal coordinate from $z$ to $(z + 2pL)$, where integer $p=0,1,2,\ldots$ is the number of round-trips and the local coordinate $z \ll L$, then using the asymptotic behavior in the short-range regime \eqref{eqn:wake short range} gives:
\begin{widetext}
\begin{equation}
    \label{eqn:wake evolution}
    \begin{aligned}
        W_{\parallel}(z + 2pL) &= \frac{2}{\pi} \sum_{mnq} \Re[Z_{\parallel,mn}(\omega_{mnq})]\cos(k_{mnq}z + 2p\psi_{mn}) \Delta\omega_\FSR\\
        &\simeq \sum_{mn} \Bigb{\cos(2p\psi_{mn})\IFT_{\cos}[Z_{\parallel,mn}](z) - \sin(2p\psi_{mn})\IFT_{\sin}[Z_{\parallel,mn}](z)},
    \end{aligned}
\end{equation}
\end{widetext}
which is a linear combination of inverse cosine and sine transforms of the impedance, with the coefficients modulated by the transverse phase shift $\psi_{mn}$.
\figref{fig:wake modes} shows the evolution of the wakefield $W_{\parallel,mn}(z + 2pL)$ of different modes over round-trips with the parameters in \tabref{tab:parameters}.
Adding all modes together gives the total wakefield evolution as shown in \figref{fig:wake evolution}.
We comment here that this decomposition of wakefield into different modes, and the analysis of their evolution over round-trips is the key difference of our work compared to previous studies.

Moreover, in contrast to the short-range wakefield $W_{\parallel}(z)$, which is strictly constrained to $z > 0$ by the principle of causality, the long-range wakefield $W_{\parallel}(z + 2pL)$ exhibits a non-zero distribution for both positive and negative $z$ (where $|z| \ll L$). This phenomenon does not violate causality; rather, it reflects the evolution of the wake into a localized wave packet after $p$ round-trips. While the initial excitation at $z \approx 0$ is a one-sided step function, the cumulative effect of mode phase shifts $\psi_{mn}$ and frequency dispersion across the discrete spectrum causes the pulse to broaden.
The degree of asymmetry between the $z < 0$ and $z > 0$ regions at $2pL$ is quantitatively governed by the term $\sum_{mn} \sin(2p\psi_{mn}) \IFT_{\sin}[Z_{\parallel,mn}](z)$, which characterizes the phase distortion of the wake as it undergoes multiple circulations within the cavity.

One may ask what if the wakefield is observed at odd multiples of the cavity length, i.e., $W_{\parallel}\bigp{z + (2p+1)L}$ where integer $p=0,1,2,\ldots$ and the local coordinate $z \ll L$.
In this case, we can also use the asymptotic behavior in the short-range regime \eqref{eqn:wake short range} to get: 
\begin{widetext}
\begin{equation}
    W_{\parallel}\bigp{z + (2p+1)L} = \frac{2}{\pi} \sum_{mnq} (-1)^q \Re[Z_{\parallel, mn}(\omega_{mnq})]\cos\bigp{k_{mnq}z + (2p+1)\psi_{mn}}\Delta\omega_\FSR
    \simeq 0,
\end{equation}
\end{widetext}
as the cavity modes are very dense in frequency domain, the Euler--Maclaurin formula ensures that odd-$q$ terms and even-$q$ terms will cause strong cancellation, leading to a negligible result compared to $W_{\parallel}(z + 2pL)$, which implies that the reversed wakefield does not affect the beam dynamics under the assumption that it does not affect the electron beam prescribed motion.
Inverse Compton scattering for example are ignored.

\begin{figure}[!t]
    \centering
    \includegraphics[width=\figwidth]{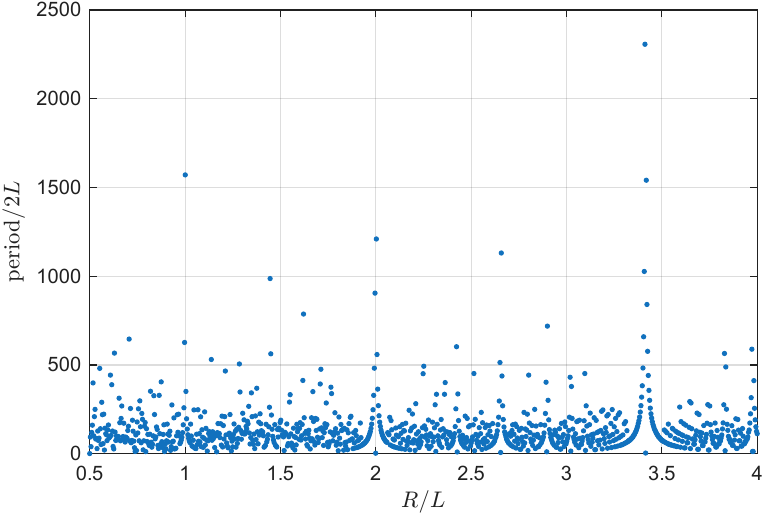}
    \caption{Periodicity of the wakefield with different cavity cavity mirror curvatures $R_1 = R_2 = R$.
    The period is calculated by tuncating the continued fraction expansion of $\psi_{00}/\pi$ with tolerance of $10^{-4}$. Lower tolerance will give a much bigger period.}
    \label{fig:wake period}
\end{figure}

Back to the discussion of the periodicity of the wakefield in the long-range regime.
The periodicity of \eqref{eqn:wake evolution} is determined by the rationality of $\psi_{mn}/\pi$.
As $\psi_{mn} = (m + n + 1)\psi_{00}$, we can only focus on the fundamental $\psi_{00}$:
if $\psi_{00}/\pi$ is a rational number, the wakefield will be strictly periodic and will repeat itself after some given number of round-trips;
if $\psi_{00}/\pi$ is an irrational number, the wakefield will be quasi-periodic, exhibiting a complex interference pattern that never exactly repeats, but may show approximate recurrences over long distances, which is determined by how well $\psi_{00}/\pi$ can be approximated by rational numbers.
\figref{fig:wake period} shows the periodicity of the wakefield with different cavity mirror curvatures $R_1 = R_2 = R$.

There are some special cases where $\psi_{00}/\pi$ is a rational number:
(i) The parallel plane cavity with $R_1 = R_2 \to \infty$, then $\psi_{00} = 0$ and the wakefield is strictly periodic with a period of $2L$;
(ii) The concentric cavity with $R_1 = R_2 = L/2$, then $\psi_{00} = \pi$ and the wakefield is strictly periodic with a period of $2L$;
(iii) The confocal cavity with $R_1 = R_2 = L$, then $\psi_{00} = \pi/2$ and the wakefield is strictly periodic with a period of $4L$.

The general case of a stable cavity with $R_1,R_2 > L/2$ usually has an irrational $\psi_{00}/\pi$, thus the wakefield is quasi-periodic.
The periodicity of the wakefield is desired to be long and not to recur too frequently, then $\psi_{00}/\pi$ needs to be close to a rational number with a big denominator.
In our case of parameters in \tabref{tab:parameters}, $\psi_{00}/\pi \approx 0.78$ is very close to $295/378$, thus the wakefield will be approximately periodic with a period of 378 round-trips.

Before ending this subsection, we may further consider the attenuation of the wakefield due to the finite finesse of the cavity.
As higher-order modes decay faster than the fundamental mode, a more precise approach is to consider the attenuation of each mode separately.
We introduce $\delta_{mn}$ as the coefficient of logarithmic fractional power loss of the HG$_{mn}$ mode per round-trip,
for simplicity here we only consider the case where $\delta_{mn}$ is independent of frequency, then the wakefield evolution can be updated as:
\begin{widetext}
\begin{equation}
    W_{\parallel}(z + 2pL) \simeq \sum_{mn} \e^{-p\delta_{mn}/2} \Bigb{\cos(2p\psi_{mn})\IFT_{\cos}[Z_{\parallel,mn}](z) - \sin(2p\psi_{mn})\IFT_{\sin}[Z_{\parallel,mn}](z)},
\end{equation}
\end{widetext}
The total cavity-loss factor $\delta_{mn}$ consists of the reflectivity $\Refl_1, \Refl_2$ of the cavity mirrors and the diffraction loss of the mode in the cavity.
\begin{equation}
    \delta_{mn} = -\ln(\Refl_1 \Refl_2) + \delta_{mn}^\diff,
\end{equation}
where the diffraction loss $\delta_{mn}^\diff$ can be calculated by integrating the mode field outside the mirror aperture
\cite{Slepian.1964.BSTJ.43.3009,Li.1965.BSTJ.44.917}.

\section{Longitudinal beam dynamics}

Having obtained the wakefield and its evolution in the cavity, we can now derive the longitudinal beam dynamics equations of a microbunch consisting of $N_\elec$ electrons.
The longitudinal coordinates of the $i$-th electron in the $j$-th turn are denoted as $(z_{i,j}, \delta_{i,j})$, where $z_{i,j}$ is the longitudinal position relative to the reference particle and $\delta_{i,j} = \gamma_{i,j}/\gamma_0 - 1$ is the relative energy deviation.

\subsection{Longitudinal dynamics equations}

To the first order, this linear coupling between the energy deviation and the longitudinal position is characterized by the longitudinal dispersion strength $R_{56}$ (also known as the momentum compaction factor in circular machines). The mapping for the longitudinal position $z_{i,j+1}$ after the $j$-th iteration or section can be expressed as:
\begin{equation}
    \label{eqn:update z}
    z_{i,j+1} = z_{i,j} + R_{56}\delta_{i,j+1}.
\end{equation}

\begin{figure}[t]
    \centering
    \includegraphics[width=0.4\textwidth, page=3]{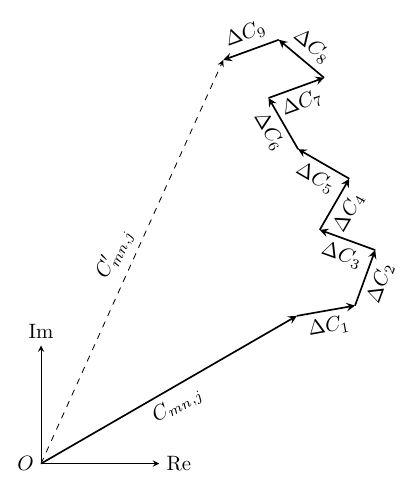}
    \caption{mode excitation $\Delta C_{mn,j}$ by a microbunch consisting of $N_\elec=9$ electrons in the laser modulator. Each electron excites the mode $\Delta C_j = \Delta C_{mn} \e^{\i k z_{i,j}}$ with a phase factor $\e^{\i k z_{i,j}}$ according to its longitudinal position $z_{i,j}$, and the total mode excitation is given by the sum of all electrons.}
    \label{fig:mode excitation}
\end{figure}

We then drive the energy deviation $\delta_{i,j+1}$ after the $j$-th turn by considering the work done by the electric field of the radiation on the electrons.
The energy of the radiation field is given by:
\begin{equation}
    W = \int\zti \dv{W}{\omega} \d\omega = \frac{2}{\pi} \sum_{mn} \int\zti P_{mn} \abs{C_{mn}}^2 \d\omega.
\end{equation}
The conservation of energy gives the relation between the change in electron energy and the work done by the electric field:
\begin{equation}
    \sum_{i=1}^{N_\elec} \gamma_{i,j} m_\elec c^2 + \frac{2}{\pi} \sum_{mn} \int\zti P_{mn} \abs{C_{mn,j}}^2 \d\omega = \mathrm{constant}, 
\end{equation}
after the microbunch passes through the undulator in the $j$-th turn, the mode amplitudes are updated from $C_{mn,j}$ to $C_{mn,j}'$, the energy deviation change of all electrons is:
\begin{equation}
    \label{eqn:Delta delta}
    \sum_{i = 1}^{N_\elec} \Delta\delta_{i,j} = -\frac{2}{\pi E_0} \sum_{mn} \int\zti\! P_{mn} \bigp{\abs{C_{mn,j}'}^2 - \abs{C_{mn,j}}^2} \d\omega,
\end{equation}
where $E_0 = \gamma_0 m_\elec c^2$ is the reference energy of the beam.
The excitation of modes $\Delta C_{mn,j}=C_{mn,j}'-C_{mn,j}$ is given by \eqref{eqn:mode excitation}.
\begin{equation}
    \Delta C_{mn,j} = \frac{e}{4P_{mn}} \sum_{i=1}^{N_\elec} \int_{-L_\und/2\bar v_z}^{L_\und/2\bar v_z} \bm{v}_{i,j} \cdot \conj{\bm{E}_{mn}}(\bm{r}_{i,j}, t) \d t.
\end{equation}
The position and velocity of the electrons can be approximated by the reference trajectory of the beam, 
\begin{subequations}
    \begin{align}
        \bm{r}_{i,j}&=\bm{r}_0+z_{i,j}\hat{\bm z},\\
        \bm{v}_{i,j}&\approx\bm{v}_0,
    \end{align}
\end{subequations}
as the electrons are tightly bunched and have small energy deviation, thus they follow the reference trajectory closely,
with the slow-varying envelope approximation: $u_{mn}(\bm{r}_{i,j}) \approx u_{mn}(\bm{r}_0)$, we have:
\begin{equation}
    \bm{E}_{mn}(\bm{r}_{i,j}) \approx \bm{E}_{mn}(\bm{r}_0) \e^{-i k z_{i,j}},
\end{equation}
then the mode excitation equation becomes:
\begin{equation}
    \label{eqn:mode excitation relation}
    \Delta C_{mn,j} = \Delta C_{mn} \sum_{i=1}^{N_\elec} \e^{\i k z_{i,j}} = \Delta C_{mn} \cdot N_\elec b_{j},
\end{equation}
where $\Delta C_{mn}$ is the mode excitation by a single reference electron at position $\bm r_0$, and $b_{j} = \sum_{i=1}^{N_\elec} \e^{\i k z_{i,j}}/N_\elec$ is the bunching factor at wavenumber $k$.
\eqref{eqn:mode excitation relation} means that the mode excitation of a single electron is a complex number with constant magnitude $\abs{\Delta C_{mn}}$ and a phase factor $\e^{\i k z_{i,j}}$ that depends on the longitudinal position of the electron, and the total mode excitation by the microbunch is given by the sum of all electrons, which is proportional to the bunching factor $b_{j}$.
The schematic of mode excitation by a microbunch in the laser modulator is shown in \figref{fig:mode excitation}.

It is worth noting that the modulation by the laser is not included in the mode excitation equation \eqref{eqn:mode excitation relation}, as the laser field is typically much stronger than the undulator radiation field ($eV_\las \gg W_\und$).
The effect of the laser modulation is separately included in the beam dynamics equations as an external modulation term, which can be added to the energy deviation update equation \eqref{eqn:Delta delta} as:
\begin{equation}
    \label{eqn:update delta}
    \delta_{i,j+1} = \delta_{i,j} + \frac{eV_\las}{E_0}\sin(k_\las z_{i,j}) + \Delta\delta_{i,j},
\end{equation}
where $V_\las$ and $k_\las$ are the effective voltage and wavenumber of the laser in the modulator, and $\Delta\delta_{i,j}$ is the energy deviation change due to the interaction with the undulator radiation field in \eqref{eqn:Delta delta}.

Moreover, when the beam is revolving in the ring turn by turn, the cavity modes also evolve due to the phase change and cavity loss.
After one ring circumference $C$ with period $T = C/\beta c$, the phase change of the modes is $\Delta\Phi_{mn} = \omega T = kC/\beta$, and the number of round-trips in the cavity is $p = cT/2L = C/2L\beta$, with attenuation factor $\e^{-\delta_{mn}/2}$ each round-trip, thus the mode amplitude is updated as:
\begin{equation}
    \label{eqn:mode evolution}
    C_{mn,j+1} = C_{mn,j}' \e^{-\delta_{mn}C/4L\beta}\e^{\i kC/\beta},
\end{equation}
which together with the mode excitation equation \eqref{eqn:mode excitation relation} gives the iteration relation of the mode amplitude in the cavity.

We can further analyze the evolution of mode amplitudes in the cavity, the change of the energy of mode $(m,n)$ due to the interaction with the bunch in the $j$-th pass is:
\begin{equation}
    \begin{aligned}
        & \abs{C_{mn,j}'}^2 - \abs{C_{mn,j}}^2 \\
        ={}& \abs{\Delta C_{mn,j}}^2 + 2\Re[\conj{\Delta C_{mn,j}} C_{mn,j}]\\
        ={}& N_\elec^2 \abs{\Delta C_{mn}}^2 \abs{b_{j}}^2 + 2N_\elec \Re[\conj{\Delta C_{mn}} \conj{b_{j}} C_{mn,j}],
    \end{aligned}
\end{equation}
the first term represents the spontaneous radiation emitted by the bunch, while the second term represents the stimulated radiation due to the existing field in the cavity.

For simplicity, here we only consider one relativistic macroparticle, i.e. $\beta = 1$ and $z_{i,j} = z_{j}$, $\delta_{i,j} = \delta_{j}$ for all $N_\elec$ electrons.
Then the iteration relation of the mode amplitude is given by:
\begin{equation}
    C_{mn,j+1} = (C_{mn,j} + N_\elec\Delta C_{mn}\e^{\i kz_{j}})\e^{-\delta_{mn}C/4L}\e^{\i kC}.
\end{equation}
At initial stage, we assume all modes are not excited, i.e. $C_{mn,0} = 0$.
Then the general term formula of the mode amplitude after $j$-th pass can be derived as:
\begin{equation}
    C_{mn,j} = N_\elec\Delta C_{mn} \sum_{p = 1}^{j} \e^{-p\delta_{mn}C/4L}\e^{\i k(z_{j-p} + pC)},
\end{equation}
then the mode energy change after the $j$-th pass can be derived as:
\begin{widetext}
\begin{equation}
    \abs{C_{mn,j}'}^2 - \abs{C_{mn,j}}^2 = N_\elec^2\abs{\Delta C_{mn}}^2 \biggp{1 + 2\sum_{p = 1}^{j} \e^{-p\delta_{mn}C/4L} \cos\bigp{k(z_{j-p} - z_{j} + pC)}},
\end{equation}
\end{widetext}
taking into the integral in \eqref{eqn:Delta delta}, the energy deviation change of the electron after the $j$-th pass can be written as wake functions:
\begin{equation}
    \Delta\delta_{j} = -\frac{N_\elec W_\und}{E_0}-\frac{N_\elec e^2}{E_0} \sum_{p = 1}^{j} W_{\parallel}(z_{j-p} - z_{j} + pC).
\end{equation}
The first term $W_\und$ represents the energy loss due to the spontaneous radiation emitted by the bunch itself; the second term $W_{\parallel}(z_{j-p} - z_{j} + pC)$ represents the energy modulation due to the stimulated radiation from the previous turns.

To ensure the phase-locking condition, the revolution time $C / \beta c$ of electron beams in the storage ring is designed to be an integer multiple of the round-trip time $2L/c$ of laser in the cavity,
which is exactly our case discussed in the previous sections.
Consequently, the mode field is reproduced at the bunch position in subsequent turns, 
We could define the scaled phase $\varPsi_{mn} := \psi_{mn}  C / \beta L$ and scaled attenuation $\Delta_{mn} := \delta_{mn} C / 4\beta L$ as shorthand notations, then the wake function and its derivative in the $p$-th turn can be written as:
\begin{widetext}
\begin{subequations}
    \begin{align}
        W_{\parallel}(z + pC) &\simeq \sum_{mn} \e^{-p\Delta_{mn}}\Bigb{\cos(p\varPsi_{mn})\IFT_{\cos}[Z_{\parallel,mn}](z) - \sin(p\varPsi_{mn})\IFT_{\sin}[Z_{\parallel,mn}](z)};\\
        \label{eqn:wake derivative evolution}
        W_{\parallel}'(z + pC) &\simeq -\sum_{mn} \e^{-p\Delta_{mn}}\Bigb{\cos(p\varPsi_{mn})\IFT_{\sin}[kZ_{\parallel,mn}](z) + \sin(p\varPsi_{mn})\IFT_{\cos}[kZ_{\parallel,mn}](z)}.
    \end{align}
\end{subequations}
\end{widetext}

\subsection{Growth rate of instability}

We can further analyze the growth rate of instability by considering the linearized beam dynamics around the equilibrium point, which is given by the fixed point of the mapping defined by \eqref{eqn:update z} and \eqref{eqn:update delta}.
Under the case of SSMB, $z_{j} \ll \lambda_\las$, 
the laser modulation can be linearized as $\sin(k_\las z) \approx k_\las z$.
Moreover, we can Taylor expand the wake function around $pC$ as:
\begin{equation}
    W_{\parallel}(z_{j-p} - z_{j} + pC) = W_{\parallel}(p C) + W_{\parallel}'(p C) (z_{j-p} - z_{j}) + \cdots,
\end{equation}
the first term $W_{\parallel}(p C)$ is the static parasitic loss term, the second term $W_{\parallel}'(p C)z_{j-p}$ characterizes the driving source and participates the collective dynamics, the third term $W_{\parallel}'(p C)z_{j}$ refers to the potential well distortion effect
\cite{Chao.1993.instability,Ng.2006.instability}.

\begin{figure}
    \centering
    \includegraphics[width=\figwidth]{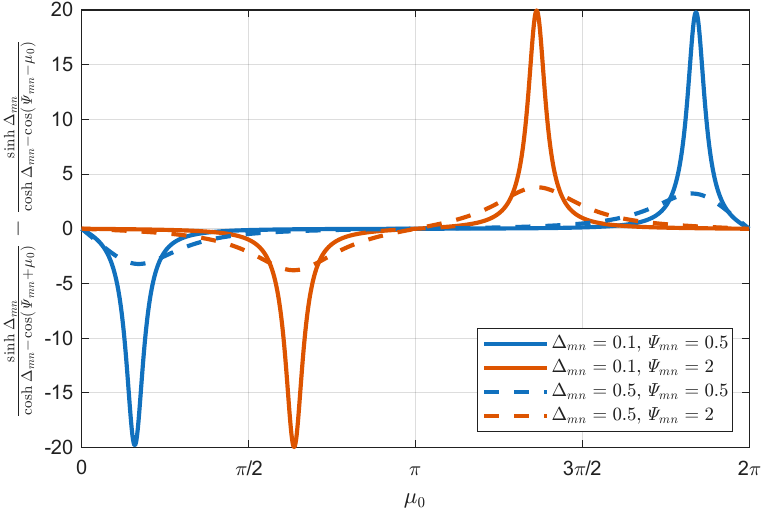}
    \caption{Resonance structure of single-mode growth rate contribution as a function of the tune $\mu_0$.
    The growth rate exhibits peaks at $\mu_0 = \pm\varPsi_{mn} + 2\pi\ell$ for integer $\ell$, with the amplitude scaling as $\bigo(1/\Delta_{mn})$ near the resonance.}
    \label{fig:bipolar}
\end{figure}

We then combine the equations \eqref{eqn:update z} and \eqref{eqn:update delta} to get the second-order difference equation for the longitudinal position of the bunch:
\begin{widetext}
\begin{equation}
    \frac{z_{j+1} - z_{j}}{R_{56}} = \frac{z_{j} - z_{j-1}}{R_{56}} + h z_{j} - \frac{N_\elec W_\und}{E_0} - \frac{N_\elec e^2}{E_0} \sum_{p=1}^{j} W_{\parallel}(p C) - \frac{N_\elec e^2}{E_0}\sum_{p=1}^{j} W_{\parallel}'(p C)(z_{j-p}-z_{j}),
\end{equation}
\end{widetext}
where $h = eV_\las k_\las / E_0$ is the energy chirp strength of the laser modulation.
The first two terms in linearized $\Delta\delta_{j}$ will be neglected as they do not depend on the current position of the electron and thus do not contribute to the growth of instability.
Then we can assume an ansatz solution of the form $z_{j} = z_0 \e^{\i j\mu}$, where $\mu$ is the complex frequency of the instability, and substitute it into the difference equation to get the dispersion relation for $\mu$:
\begin{equation}
    \frac{(\e^{\i\mu} - 1)^2}{\e^{\i\mu}} = hR_{56} - \frac{N_\elec e^2R_{56}}{E_0}\sum_{p=1}^{\infty} W_{\parallel}'(p C)(\e^{-\i p\mu}-1).
\end{equation}
here we have taken the limit of $j \to \infty$ to get the asymptotic growth rate of instability, and the summation is taken to infinity as the wakefield decays with $p$.
Under small frequency limit $\abs{\mu} \ll 1$, we can take $(\e^{\i\mu} - 1)^2/\e^{\i\mu} \approx -\mu^2$.
If there is no wakefield, i.e. $W_{\parallel}(z) \equiv 0$, then the dispersion relation reduces to $\mu_0^2 = -h R_{56}$, which is the synchrotron oscillation frequency of the beam in the presence of the laser modulation.
The frequency shift due to the wakefield $\Delta\mu = \mu - \mu_0$ can be calculated by treating the wakefield term as a perturbation where $\abs{\Delta\mu} \ll \abs{\mu_0}$, which gives:
\begin{equation}
    \Delta\mu = \frac{N_\elec e^2 R_{56}}{2\mu_0 E_0}\sum_{p=1}^{\infty} W_{\parallel}'(p C)(\e^{-\i p\mu_0}-1).
\end{equation}
The growth rate of the instability is given by the imaginary part of $\Delta\mu$, which can be calculated as:
\begin{equation}
    \label{eqn:growth rate}
    \frac{1}{\tau} = -\frac{\Im[\Delta\mu]}{T} = \frac{N_\elec e^2c R_{56}}{2\mu_0 E_0 C}\sum_{p=1}^{\infty} W_{\parallel}'(p C)\sin(p\mu_0),
\end{equation}
where $T = C/c$ is the round-trip time of the beam in the storage ring.
The series related to the wakefield derivative can be further simplified by substituting \eqref{eqn:wake derivative evolution} into it, which gives:
\begin{widetext}
\begin{equation}
    \label{eqn:growth rate series}
    \begin{aligned}
        \ifispreprint
        &\sum_{p=1}^{\infty} W_{\parallel}'(p C)\sin(p\mu_0) \\
        \else
        \sum_{p=1}^{\infty} W_{\parallel}'(p C)\sin(p\mu_0)
        \fi
        &= -\sum_{p=1}^{\infty} \biggb{\sin(p\mu_0) \sum_{mn} \e^{-p\Delta_{mn}} \sin(p\varPsi_{mn})\IFT_{\cos}[kZ_{\parallel,mn}](0)}\\
        &= \frac{1}{2} \sum_{mn} \biggb{\IFT_{\cos}[kZ_{\parallel,mn}](0) \sum_{p=1}^{\infty} \e^{-p\Delta_{mn}} \bigb{\cos\bigp{p(\varPsi_{mn} + \mu_0)} - \cos\bigp{p(\varPsi_{mn} - \mu_0)}}}\\
        &= \frac{1}{4} \sum_{mn} \biggb{\frac{\sinh\Delta_{mn}}{\cosh\Delta_{mn} - \cos(\varPsi_{mn} + \mu_0)} - \frac{\sinh\Delta_{mn}}{\cosh\Delta_{mn} - \cos(\varPsi_{mn} - \mu_0)}} \IFT_{\cos}[kZ_{\parallel,mn}](0).
    \end{aligned}
\end{equation}
\end{widetext}
When $\Delta_{mn} \ll 1$, the term in the bracket of a single mode attains a sharp maximum value of $\bigo(1/\Delta_{mn})$ at the resonant position $\mu_0 = -\varPsi_{mn} + 2\pi\ell$ and a corresponding minimum at $\mu_0 = \varPsi_{mn} + 2\pi\ell$, with a peak width of $\bigo(\Delta_{mn})$, which is shown in \figref{fig:bipolar}.
The superposition of all modes gives the total growth rate, which exhibits a complex resonance structure with multiple peaks and valleys as a function of the tune $\mu_0$.

\begin{figure*}[t]
    \centering
    \includegraphics[width=\linewidth]{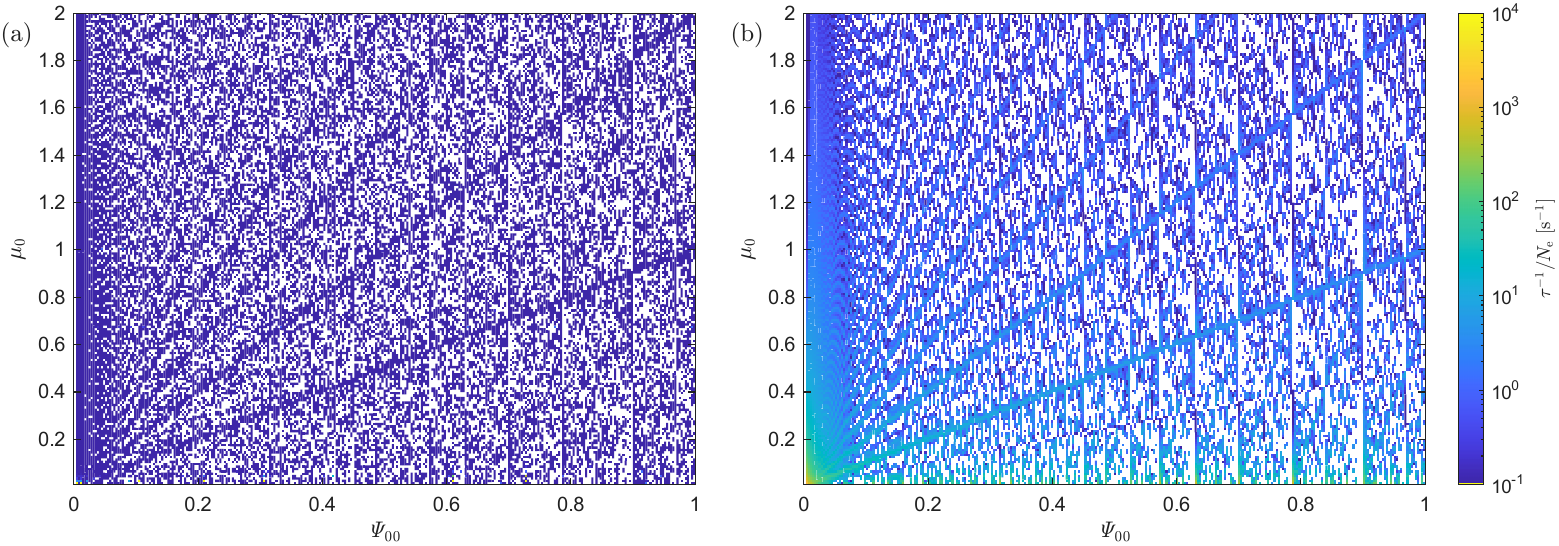}
    \caption{Growth rate of the instability for different phase shift $\varPsi_{00}$ and tune $\mu_0$ under (a) low cavity loss $\Delta_{mn} \equiv 10^{-5}$ for all modes and (b) high cavity loss $\Delta_{mn} \equiv 10^{-2}$ for all modes.
    White regions represent the stable region with negative growth rate, while colored regions represent the unstable region with positive growth rate, and the color intensity represents the magnitude of the growth rate.
    Both (a) and (b) share the same colorbar for the growth rate.}
    \label{fig:growth rate}
\end{figure*}

The dependence of the growth rate \eqref{eqn:growth rate} on the fundamental phase shift $\varPsi_{00}$ and the tune $\mu_0$ is shown in \figref{fig:growth rate} for both low and high cavity loss cases.
The following features characterize the system's stability: 
(i) While a smaller $\Delta_{mn}$ leads to a sharper resonance with higher peak growth rates as the wakefield can persist over more turns, it also restricts the most unstable regions to narrower bands.
(ii) The map reveals distinct instability lines following the relation $\mu_0 = \varPsi_{mn} = (m + n + 1)\varPsi_{00}$, corresponding to the parameter sets reaching the minimum. These features demonstrate that each transverse mode acts as an independent resonant driver, but modulated by its unique Gouy phase shift.
(iii) Consequently, the total growth rate is not a simple superposition but an intricate resonance pattern. This complexity arises from the coherent interference of hundreds of cavity modes, leading to a distribution that is highly sensitive to small perturbations in the cavity geometry or the storage ring working point.

The physical mechanism of the longitudinal instability investigated here is analogous to the classical Robinson instability observed in conventional RF cavities.
In both cases, the instability arises from the cumulative interaction between the electron bunch and a long-range electromagnetic field trapped within a high-$Q$ resonator.
However, several distinct features set the SSMB laser modulator system apart:
First, while the Robinson instability is typically driven by a single dominant longitudinal mode of an RF cavity, the instability in an SSMB laser modulator is a multi-mode collective effect. It involves the coherent superposition of a complete set of transverse Hermite--Gaussian eigenmodes.
Second, the phase evolution of these modes is not solely determined by frequency detuning, but is significantly modulated by the Gouy phase shift of optical modes. This results in a more complex resonance structure, where each transverse mode contributes its own phase slip to the total wakefield.
Consequently, the stability of the beam is governed by the derivative of the accumulated multi-mode wakefield at the bunch position, leading to the intricate resonance patterns observed in our growth rate maps in \figref{fig:growth rate}.

Finally, it should be emphasized that the convergence of high-order cavity modes is relatively slow, with the fundamental mode accounting for only a fraction of the total contribution. Consequently, high-order modes exert a significant influence on the instability.
In the practical design of laser cavities for SSMB, various techniques such as the implementation of D-shaped mirrors are typically employed to accelerate the attenuation of these high-order modes
\cite{Liu.2024.RSI.95.103004}.
Such measures serve as an effective means to suppress the instabilities induced by high-order transverse modes.
Futher details in this respect are expected to be discussed in our later work.

\section{Numerical results}

As an example, we simulate the longitudinal beam dynamics in an SSMB storage ring with parameters listed in \tabref{tab:parameters}
\cite{Deng.2024.PhD}.
Here we consider a constant cavity loss $\Delta_{mn} \equiv \num{1e-5}$ for all modes for simplicity.

The calculations of the wakefield and its evolution in the cavity are performed before the beam dynamics simulation, and the results are stored in a look-up table for efficient interpolation during the beam dynamics simulation.

\begin{table}[t]
    \centering
    \caption{Parameters for an SSMB example.}
    \label{tab:parameters}
    \begin{ruledtabular}
    \begin{tabular}{lll}
        Parameter & Value & Description \\
        \hline
        $E_0$ & \SI{250}{MeV} & Beam energy \\
        $\lambda_\rad$ & \SI{1064}{nm} & Radiation wavelength \\
        $\lambda_\las$ & \SI{1064}{nm} & Modulation laser wavelength \\
        $K$ & 1.76 & Undulator parameter \\
        $h$ & \SI{5000}{m^{-1}} & Energy chirp strength \\
        $\lambda_\und$ & \SI{20}{cm} & Undulator period \\
        $N_\und$ & 10 & Number of undulator periods \\
        $L_\und=N_\und\lambda_\und$ & \SI{2}{m} & Undulator length \\
        $L$ & \SI{4}{m} & Cavity length \\
        $R_1 = R_2$ & \SI{2.26}{m} & Radii of cavity mirror curvature \\
        $\Refl_1 = \Refl_2$ & \num{0.99999} & Reflectivity of cavity mirrors \\
        $\zR$ & \SI{0.72}{m} & Rayleigh length \\
        $C$ & \SI{48}{m} & Storage ring circumference \\
        $R_{56}$ & \SI{-10}{\micro m} & Storage ring dispersion \\
    \end{tabular}
    \end{ruledtabular}
\end{table}

\begin{figure}[t]
    \centering
    \includegraphics[width=\figwidth]{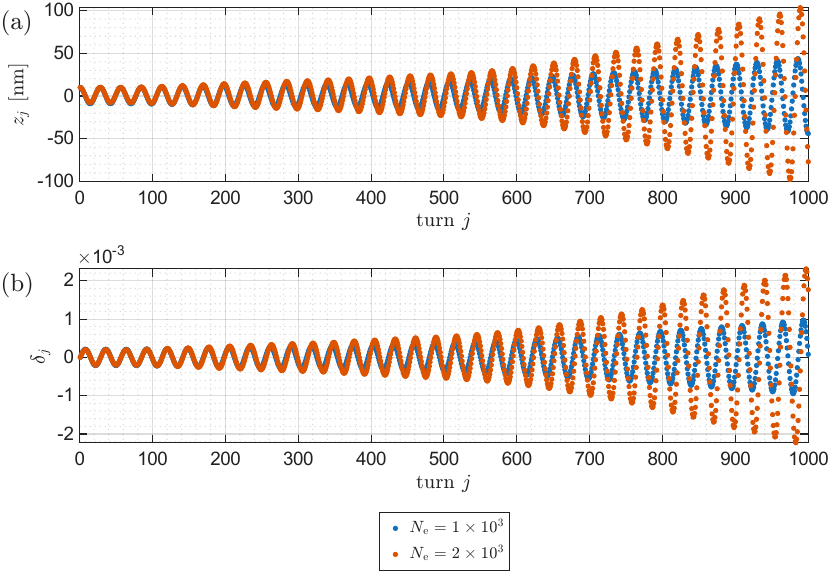}
    \caption{Evolution of the longitudinal phase space coordinates $(z_j, \delta_j)$ of the electron beam over multi-turn in the storage ring.
    (a) $z_j$ versus turn number $j$; (b) $\delta_j$ versus turn number $j$.}
    \label{fig:evolution zdelta}
\end{figure}

The evolution of the longitudinal phase space coordinates $(z_j, \delta_j)$ of the macroparticle with electron number $N_\elec = \num{1e3}$ over multi-turn in the storage ring is shown in \figref{fig:evolution zdelta}.
The initial condition of the macroparticle is set to be $z_0=\SI{10}{nm}$ and $\delta_0=0$.

At the beginning, the macroparticle undergoes synchrotron oscillations with the frequency of $\mu_0 \approx 0.22$, the macroparticle are mainly modulated by the laser with amplitude of $eV_\las/E_0$.
As the radiation field builds up in the cavity over multiple turns, the energy modulation of the electron beam grows due to the interaction with the coherent undulator radiation field, which leads to the growth of energy center oscillation amplitude in the longitudinal phase space. During this stage, the oscillation amplitude exhibits a distinct exponential growth trend with the number of turns. 
Eventually, the energy modulation can become large enough to cause significant distortion of the beam distribution and even beam loss if the energy spread exceeds half the bucket height $\delta_{1/2}$.

\begin{figure}[th]
    \centering
    \includegraphics[width=\figwidth]{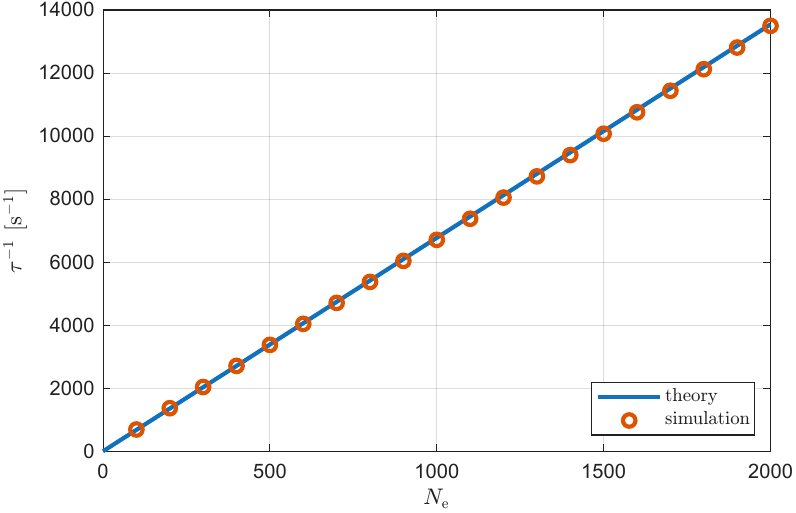}
    \caption{Growth rate as a function of electron number $N_\elec$. Blue line refers to the theoretical growth rate calculated by \eqref{eqn:growth rate}, and orange dots refer to the growth rate fitted from the simulation results.}
    \label{fig:growth rate-Ne}
\end{figure}

Under small electron number, the wakefield is much smaller than the laser ($N_\elec W_\und \ll eV_\las$).
The longitudinal coordinate $z_j$ of the macroparticle could be well fitted by a sinusoidal function of turn number $j$:
\begin{equation}
    z_j = z_0 \e^{jT/\tau} \cos(\mu_0 j),
\end{equation}
where $\tau^{-1}$ is the growth rate of the instability.
\figref{fig:growth rate-Ne} shows the growth rate as a function of electron number $N_\elec$, and we find good agreement between the growth rate fitted from the simulation results and the theoretical growth rate calculated by \eqref{eqn:growth rate}.

As the electron number increases, the growth rate of the instability is eventually serious enough to cause beam loss within a few thousand turns, thus it is crucial to consider the effect of this instability in the design of SSMB storage rings.
Nevertheless, the growth rate can be effectively reduced by optimizing the cavity geometry to increase the attenuation of high-order modes, or by tuning the storage ring working point to avoid the resonance peaks in the growth rate map shown in \figref{fig:growth rate}.

\section{Summary and outlook}

In this paper, we have established a comprehensive theoretical and numerical framework to investigate the longitudinal beam instabilities driven by coherent undulator radiation within the laser modulator of an SSMB storage ring.
While previous studies relied on rigid wakefield models, our approach employs a rigorous cavity mode decomposition technique. By expanding the radiation field into a complete set of Hermite-Gaussian modes, we have successfully captured the evolution of the wakefields as they circulate within the optical cavity.

The core of our work involved deriving the mode excitation equations based on first principles, treating the electron beam as a current source. This allowed us to calculate the specific mode decomposition of the undulator radiation spectrum and trace its turn-by-turn evolution. Our analysis revealed several critical insights:

Instability mechanism:
We demonstrated that the interaction between the electron bunch and the accumulated radiation field leads to a cumulative energy modulation.
This process is analogous to Robinson instability but is uniquely characterized by the discrete resonance conditions of the laser cavity and the transverse phase shifts (Gouy phase) of the various cavity modes.
Through both analytical derivations and numerical simulations, we characterized the growth rate of the instability.

Role of high-order modes:
An important finding of this study is the non-negligible contribution of high-order transverse modes.
Unlike free-space propagation, the cavity environment traps these modes, and their slower convergence necessitates their inclusion for accurate instability threshold predictions.
Our results suggest that the resonance structure of the growth rate is highly sensitive to the phase shifts of these high-order modes.

Parametric sensitivity:
The simulation results indicate that the turn number at which beam loss occurs is inversely proportional to the beam charge and increases with the number of undulator periods. Furthermore, the cavity loss coefficients play a dual role: while smaller losses allow for higher average power, they also broaden the unstable domains by allowing wakefields to persist over a greater number of turns.

From a design perspective, our work provides a quantitative tool for optimizing the performance of SSMB storage rings. To mitigate the adverse effects of these instabilities, we propose the following strategies:

Mode control:
The implementation of specialized cavity geometries, such as the use of D-shaped mirrors or aperture-limiting elements, can be employed to selectively increase the attenuation of high-order modes without significantly degrading the fundamental mode.

Operational tuning:
By carefully choosing the cavity mirror curvature and the storage ring tune, operators can steer the system away from the sharp resonance peaks identified in our growth rate maps, thereby extending the beam lifetime.

The framework developed in this study will serve as a foundation for several promising avenues of future research.
Firstly, the current model assumes a simplified cavity geometry; extending the mode decomposition method to include complex asymmetric 4-mirror cavity configurations would enhance its applicability to advanced light source designs.
Secondly, while this paper focuses on longitudinal dynamics, a unified 6D phase-space analysis incorporating transverse-longitudinal coupling is essential to capture the full picture of beam-radiation interaction.

Finally, the transition from single-bunch to multi-bunch multi-turn scenarios represents a vital next step. In high-repetition-rate SSMB rings, the long-range wakefields from preceding bunches will inevitably influence trailing bunches, potentially leading to coupled-bunch instabilities. Expanding our mode evolution equations to account for these collective effects will be instrumental in the realization of next-generation, high-average-power coherent radiation sources.

\begin{acknowledgments}
The authors would like to thank Prof. Alexander Chao, Prof. Fabian Zomer, and Prof. Cheng-Ying Tsai for their helpful discussions and suggestions on this work.
This work is supported by
the National Key Research and Development Program of China (Grant No. 2022YFA1603403),
the Beijing Outstanding Young Scientist Program (No. JWZQ20240101006),
the National Natural Science Foundation of China (NSFC Grants No. 12522512, 124B2104),
and
Tsinghua University Dushi Program.
\end{acknowledgments}

\bibliography{ref}

\end{document}

%% file: mathdef.tex
\let\Re\relax
\let\Im\relax
\DeclareMathOperator{\Re}{Re}  
\DeclareMathOperator{\Im}{Im}  
\DeclareMathOperator{\FT}{\mathcal{F}}  
\newcommand{\IFT}{\FT^{-1}}  

\newcommand{\e}{\mathop{}\!\mathrm{e}}
\renewcommand{\i}{\mathrm{i}}

\newcommand{\nd}{\mathrm{d}}
\renewcommand{\d}{\mathop{}\!\nd}
\newcommand{\p}{\partial}
\newcommand{\bigo}{\mathcal{O}}

\newcommand*{\dv}[3][{}]{\frac{\nd^{#1}#2}{\nd{#3}^{#1}}}
\newcommand*{\dw}[3]{\frac{\nd^2{#1}}{\nd{#2}\nd{#3}}}

\newcommand*{\pv}[3][{}]{\frac{\p^{#1}#2}{\p{#3}^{#1}}}

\newcommand{\zti}{_0^{+\infty}}
\newcommand{\iti}{_{-\infty}^{+\infty}}

\newcommand{\abs}[1]{\lvert#1\rvert}     

\newcommand{\biggabs}[1]{\biggl\lvert#1\biggr\rvert}
\newcommand{\bigp}[1]{\bigl(#1\bigr)}       
\newcommand{\Bigp}[1]{\Bigl(#1\Bigr)}
\newcommand{\biggp}[1]{\biggl(#1\biggr)}
\newcommand{\bigb}[1]{\bigl[#1\bigr]}       
\newcommand{\Bigb}[1]{\Bigl[#1\Bigr]}
\newcommand{\biggb}[1]{\biggl[#1\biggr]}

\newcommand*{\conj}[1]{#1^*}  

\newcommand{\elec}{\mathrm{e}}

\newcommand{\und}{\mathrm{u}}

\newcommand{\rad}{\mathrm{r}}

\newcommand{\las}{\mathrm{L}}
\newcommand{\zR}{z_{\mathrm{R}}}  
\newcommand{\zRlas}{z_{\mathrm{R},\las}}

\newcommand{\FSR}{\mathrm{FSR}}

\newcommand{\diff}{\mathrm{diff}}

\newcommand*{\Refl}{\mathcal{R}}  

\newcommand*{\figref}[1]{FIG.~\ref{#1}}
\newcommand*{\tabref}[1]{Table~\ref{#1}}